\documentclass[12pt]{iopart}

\usepackage{graphicx}

\begin{document}

\title[An efficient source of continuous variable polarisation entanglement]{An efficient source of continuous variable polarisation entanglement}

\author{Ruifang~Dong$^1$, Joel~Heersink$^1$, Jun--ichi Yoshikawa$^2$, Oliver~Gl\"ockl$^1$, Ulrik~L.~Andersen$^{1,3}$, Gerd~Leuchs$^1$, 
}

\address{$^1$ Institut f\"ur Optik, Information und Photonik, Max--Planck Forschungsgruppe,
Universit\"at Erlangen--N\"urnberg, G\"unther--Scharowsky--Stra{\ss}e 1, 91058 Erlangen, Germany}
\address{$^2$ Department of Applied Physics, School of Engineering, University of Tokyo, Japan}
\address{$^3$ Department of Physics, Technical University of Denmark, Building 309, 2800 Lyngby, Denmark}
\ead{rdong@optik.uni-erlangen.de}

\begin{abstract}
We have experimentally demonstrated the efficient creation of highly entangled bipartite continuous variable polarisation states. Exploiting an optimised scheme for the production of squeezing using the Kerr non--linearity of a glass fibre we generated polarisation squeezed pulses with a mean classical excitation in $\hat{S}_3$. Polarisation entanglement was generated by interfering two independent polarisation squeezed fields on a symmetric beam splitter. The resultant beams exhibit strong quantum noise correlations in the dark $\hat{S}_1-\hat{S}_2$ polarisation plane. To verify entanglement generation, we characterised the quantum correlations of the system for two different sets of conjugate Stokes parameters. The quantum correlations along the squeezed and the anti--squeezed Stokes parameters were observed to be -4.1$\pm$0.3~dB and -2.6$\pm$0.3~dB below the shot noise level respectively. The degree of correlations was found to depend critically on the beam splitting ratio of the entangling beam splitter. Carrying out measurements on a different set of conjugate Stokes parameters, correlations of -3.6$\pm$0.3~dB and -3.4$\pm$0.3~dB have been observed. This result is more robust against asymmetries in the entangling beam splitter, even in the presence of excess noise.
\end{abstract}


\maketitle

\section{Introduction}

Entanglement has enjoyed a special place in physics ever since its inception in 1935~\cite{einstein35.pr}. In this Gedankenexperiment, a pair of states was postulated in which the ability to infer the value of an observable of a second system based on observations of the first system is better than quantum mechanics seems to allow. Such counterintuitive states exhibit correlations of a quantum nature and were first measured in the discrete variable regime in 1950~\cite{wu50.pr}. Subsequent theoretical and experimental investigations of such correlated states has in a large part been responsible for the wealth of phenomena and protocols in modern quantum optics. Of particular interest to this paper is the extension and demonstration of these ideas in the continuous variable regime. The first such experiment was carried out by Ou~\textit{et al.}~\cite{ou92.prl}. These and numerous further experiments employ continuous quantum observables such as the amplitude and phase quadratures of the electromagnetic field. Entanglement between these variables is then analogous to the position and momentum of the original EPR Gedankenexperiment~\cite{einstein35.pr}.

In the case of intense light fields, its polarisation state can also be described by a set of continuous variables which can be entangled. The advantage of the polarisation variables in quantum optics, typically described by the Stokes operators~\cite{jauch55.book}, over the quadrature variables is their ease of detection. The Stokes operators can be directly detected taking advantage of an internal phase reference and thus cumbersome measurements with local oscillators are unnecessary~\cite{korolkova02.pra}.

The early experiments on polarisation squeezing used continuous wave light and parametric processes~\cite{grangier87.prl, hald00.jomo, bowen02a.prl}. Since these experiments, polarisation squeezing has been experimentally demonstrated using silica fibres~\cite{heersink03.pra, heersink05.ol} and cold atomic samples~\cite{josse03.prl}. Such non--classical states are the building blocks for continuous variable polarisation entanglement utilising only passive elements such as beam splitters. Extending the ideas of quadrature entanglement~\cite{ou92.prl, silberhorn01.prl}, polarisation entanglement was suggested~\cite{korolkova02.pra}. The first realisation of such quantum correlated states was shown by appropriately transforming a quadrature entangled state~\cite{bowen02b.prl} into a polarisation entangled state. A further solution was demonstrated in Ref.~\cite{gloeckl03.job} where a single polarisation squeezed input was used for the entanglement creation. In this paper we build upon the latter methods and use two polarisation squeezed input states to generate polarisation entanglement. We develop and characterise an efficient source of pulsed polarisation entanglement using an optimised fibre based polarisation squeezing setup~\cite{heersink05.ol}.

\section{Polarisation entanglement}

The quantum polarisation state of an intense light field can be succinctly described by the quantum Stokes operators~\cite{jauch55.book, robson74.book, chirkin93.qe}. These are derived by analogy to the classical Stokes parameters~\cite{stokes52.tcps}. If $\hat{a}_{x/y}$ and $\hat{a}^{\dagger}_{x/y}$ denote the photon annihilation and creation operators of two orthogonal polarisation modes $x$ and $y$, and $\hat{n}_x$ and $\hat{n}_y$ are the photon number operators of these modes, the quantum Stokes operators read as follows
\begin{eqnarray}
\hat{S}_{0} &=& \hat{a}^{\dagger}_{x} \hat{a}_{x} + \hat{a}^{\dagger}_{y}
\hat{a}_{y} = \hat{n}_x + \hat{n}_y, \nonumber \\
\hat{S}_{1} &=& \hat{a}^{\dagger}_{x} \hat{a}_{x} - \hat{a}^{\dagger}_{y}
\hat{a}_{y} = \hat{n}_x - \hat{n}_y, \nonumber \\
\hat{S}_{2} &=& \hat{a}^{\dagger}_{x} \hat{a}_{y} + \hat{a}^{\dagger}_{y}
\hat{a}_{x}, \nonumber \\
\hat{S}_{3} &=& i(\hat{a}^{\dagger}_{y} \hat{a}_{x} - \hat{a}^{\dagger}_{x}
\hat{a}_{y}).
\end{eqnarray}
The operators $\hat{S}_1$, $\hat{S}_2$, and $\hat{S}_3$ follow the operator valued commutation relation of the SU(2) Lie algebra
\begin{equation}\label{eq_commutator}
[ \hat{S}_k,\hat{S}_l]  = 2i\varepsilon_{klm}\hat{S}_m.
\end{equation}
The $\hat{S}_0$ operator, which represents the total intensity, commutes with all other Stokes operators. This formalism corresponds to the well known derivation of Schwinger \cite{schwinger65.book}. It gives rise to a set of three state dependent Heisenberg-type uncertainty relations
\begin{eqnarray}
\Delta^2\hat{S}_1 \Delta^2\hat{S}_2 \geq |\langle\hat{S}_3\rangle|^{2}, \nonumber\\
\Delta^2\hat{S}_3 \Delta^2\hat{S}_1 \geq |\langle\hat{S}_2\rangle|^{2}, \nonumber\\
\Delta^2\hat{S}_2 \Delta^2\hat{S}_3 \geq |\langle\hat{S}_1\rangle|^{2}.   \label{eqn:stokes_uncertainty}
\end{eqnarray}
Thus a state which obeys
\begin{equation}
\Delta^2\hat{S}_k < |\langle\hat{S}_l\rangle| < \Delta^2\hat{S}_m, \qquad k\neq l \neq m,
\end{equation}
is a polarisation squeezed state (see Refs.~\cite{korolkova02.pra, heersink05.ol} and references therein). In the experiments presented in this paper we consider a fully circularly polarized state, i.e.\ $\langle\hat{S}_3\rangle \neq 0$. The Stokes parameters orthogonal to $\hat S_3$ are given by
\begin{equation}
\hat{S}(\theta) = \cos(\theta) \hat{S}_1 + \sin(\theta) \hat{S}_2,\label{pol_sq_darkplane}
\end{equation}
all of which have the property $\langle\hat{S}(\theta)\rangle = 0$. Physically speaking these parameters are dark. This plane contains an infinite family of maximally conjugate Stokes parameters, generally given by $\hat{S}(\theta)$ and $\hat{S}(\theta+\pi/2)$ which obey the uncertainty relation
\begin{equation}
\Delta^2\hat{S}(\theta) \Delta^2\hat{S}(\theta+\pi/2) \geq |\langle\hat{S}_3\rangle|^2.
\end{equation}
Thus the values of $\hat{S}(\theta)$ and $\hat{S}(\theta+\pi/2)$ cannot be simultaneously determined with arbitrary accuracy. That is, the variance $\Delta^2\hat{S}(j)$ of $\hat{S}(j)$ cannot vanish for $j=\theta$ and $j=\theta+\pi/2$ simultaneously. The variance $V_\mathrm{sq}$ of a polarisation squeezed state is minimised for a particular angle $\theta_\mathrm{sq}$, i. e. $\hat S(\theta_\mathrm{sq})=\hat S_\mathrm{sq}$ is squeezed, the variance $V_\mathrm{asq}$ of the corresponding conjugate parameter $\hat S(\theta_\mathrm{sq}+\pi/2)=\hat S_\mathrm{asq}$ is anti--squeezed.  

Such polarisation squeezed states can be used for the generation of polarisation entanglement. Polarisation entanglement can be quantified and characterised in many ways, depending on the application and precise system under consideration~\cite{eisert03.ijqi}. However, the most common measures used for the experimental demonstration of continuous variable entanglement are the EPR criterion~\cite{reid89.pra} and the non--separability criterion~\cite{duan00_simon00}. Although originally derived for the characterisation of quadrature entanglement these criteria have equivalents in the polarisation variables~\cite{korolkova02.pra}. In general, let us consider a pair of Stokes parameters $\hat S(\theta)$ and $\hat S(\theta+\pi/2)$ which form a conjugate pair. Consider a composite quantum system that consists of two modes which we label C and D. For such a system, following from basic commutation relations (\ref{eq_commutator}) the occurance of simultaneous quantum correlations of the type $\hat S_\mathrm{C}(\theta)+\hat S_\mathrm{D}(\theta)\rightarrow0$ and $\hat S_\mathrm{C}(\theta+\pi/2)-\hat S_\mathrm{D}(\theta+\pi/2)\rightarrow0$ are a signature for entanglement. Another combination between the two modes that can exhibit strong quantum correlations and hence shows entanglement is given by $\hat S_\mathrm{C}(\theta)+\hat S_\mathrm{D}(\theta+\pi/2)\rightarrow0$ and $\hat S_\mathrm{C}(\theta+\pi/2)+\hat S_\mathrm{D}(\theta)\rightarrow0$. Note that the two types of correlations are equivalent, as the latter case can be achieved by performing a linear operation of the type $\hat S_\mathrm{D}(\theta)\rightarrow -\hat S_\mathrm{D}(\theta+\pi/2)$ and $\hat S_\mathrm{D}(\theta+\pi/2)\rightarrow \hat S_\mathrm{D}(\theta)$ on one of the entangled modes. This corresponds to a rotation of the polarisation by 45$^\circ$ using a $\lambda/2$--wave plate, i.e. a rotation in phase space by 90$^\circ$.

Using these considerations a state is called polarisation entangled if

\begin{eqnarray}
\sqrt{\Delta^2(\hat{S}_\mathrm{C}(\theta)+\hat{S}_\mathrm{D}(\theta)) \cdot \Delta^2(\hat{S}_\mathrm{C}(\theta+\pi/2)-\hat{S}_\mathrm{D}(\theta+\pi/2) )} &<& ( \vert \langle \hat{S}_\mathrm{3,C}\rangle \vert + \vert \langle \hat{S}_\mathrm{3,D}\rangle \vert ),
\label{PeresHorodecki_productform}
\end{eqnarray}
or
\begin{eqnarray}
\sqrt{\Delta^2(\hat{S}_\mathrm{C}(\theta)+\hat{S}_\mathrm{D}(\theta+\pi/2)) \cdot \Delta^2(\hat{S}_\mathrm{C}(\theta+\pi/2)+\hat{S}_\mathrm{D}(\theta) )} &<& ( \vert \langle \hat{S}_\mathrm{3,C}\rangle \vert + \vert \langle \hat{S}_\mathrm{3,D}\rangle \vert ),
\label{PeresHorodecki_productform_2}
\end{eqnarray}
The non--separability criterion in product form \cite{nonsep_product} to witness continuous variable entanglement is more general than the sum criterion by Duan and Simon~\cite{duan00_simon00}. In addition, for symmetric states the product of the correlations can be used to quantify the amount of entanglement in terms of the entanglement of formation (EOF) \cite{Giedke_EOF}.
The sum (difference) variances of the Stokes operators on the left hand side of (\ref{PeresHorodecki_productform}) quantify the quantum correlations between the subsystems C and D in the respective conjugate variables. The right hand side provides the reference quantum noise limit according to the uncertainty relation Eq.~(\ref{eqn:stokes_uncertainty}), see also~\cite{korolkova02.pra, bowen02b.prl}. A state which is non--separable according to Eq.~(\ref{PeresHorodecki_productform}) can be generated e.g.\ by the interference of one or two polarisation squeezed light fields on a beam splitter. Here we consider the case of two independent, but equally polarisation squeezed beams A and B as seen in Fig.~\ref{polentsetup}. This is in analogy to common knowledge of the generation of quadrature entanglement, see for example Refs.~\cite{ent_bs}. We expect statistically identical output beams C and D and the correlations in both conjugate variables should be equal. This is in contrast to our previous resource efficient experiment using only one polarisation squeezed input~\cite{gloeckl03.job}.

The pair of polarisation squeezed beams A and B are described by their respective Stokes operators along the squeezing and the anti--squeezing direction in the dark plane: $\hat S_\mathrm{A,B}(\theta_\mathrm{sq})$ and $\hat S_\mathrm{A,B}(\theta_\mathrm{sq}+\pi/2)$. The corresponding variances are denoted $V_\mathrm{sq}$ and $V_\mathrm{asq}$. Via the interference of these two polarisation squeezed fields on a beam splitter with a relative optical phase of $\pi/2$, polarisation entanglement can be generated. The entangled modes are labeled by C and D. The input output relations for the Stokes parameters along the squeezing and the anti--squeezing direction in the dark plane for a beam splitter with reflectivity $\sqrt{R}$ and transmittivity $\sqrt{T}$ are given by

\begin{eqnarray}\label{eq_pol_ent_bs}
\hat S_\mathrm{C}(\theta_\mathrm{sq})&=&T\hat S_\mathrm{A}(\theta_\mathrm{sq})+R\hat S_\mathrm{B}(\theta_\mathrm{sq})\nonumber \\&&+
						\sqrt{RT}\hat S_\mathrm{A}(\theta_\mathrm{sq}+\pi/2)-\sqrt{RT}\hat S_\mathrm{B}(\theta_\mathrm{sq}+\pi/2),\nonumber \\ 
\hat S_\mathrm{D}(\theta_\mathrm{sq})&=&R\hat S_\mathrm{A}(\theta_\mathrm{sq})+T\hat S_\mathrm{B}(\theta_\mathrm{sq})\nonumber \\&&-
						\sqrt{RT}\hat S_\mathrm{A}(\theta_\mathrm{sq}+\pi/2)+\sqrt{RT}\hat S_\mathrm{B}(\theta_\mathrm{sq}+\pi/2),\nonumber \\
\hat S_\mathrm{C}(\theta_\mathrm{sq}+\pi/2)&=&T\hat S_\mathrm{A}(\theta_\mathrm{sq}+\pi/2)+R\hat S_\mathrm{B}(\theta_\mathrm{sq}+\pi/2)\nonumber\\
&&-\sqrt{RT}\hat S_\mathrm{A}(\theta_\mathrm{sq})+\sqrt{RT}\hat S_\mathrm{B}(\theta_\mathrm{sq}),\nonumber \\
\hat S_\mathrm{D}(\theta_\mathrm{sq}+\pi/2)&=&R\hat S_\mathrm{A}(\theta_\mathrm{sq}+\pi/2)+T\hat S_\mathrm{B}(\theta_\mathrm{sq}+\pi/2)\nonumber \\&&+\sqrt{RT}\hat S_\mathrm{A}(\theta_\mathrm{sq})-\sqrt{RT}\hat S_\mathrm{B}(\theta_\mathrm{sq}).										
\end{eqnarray}

As can be seen from these equations, the polarisation states of the two output modes exhibit strong quantum correlations along the initially squeezed $\hat S(\theta_\mathrm{sq})$ and anti--squeezed $\hat S(\theta_\mathrm{sq}+\pi/2)$ directions. For a symmetric beam splitting ratio the quantum correlations of the type $\hat S_\mathrm{C}(\theta_\mathrm{sq})+\hat S_\mathrm{D}(\theta_\mathrm{sq})\rightarrow0$ and $\hat S_\mathrm{C}(\theta_\mathrm{sq}+\pi/2)-\hat S_\mathrm{D}(\theta_\mathrm{sq}+\pi/2)\rightarrow0$ are optimised, limited only by the amount of polarisation squeezing in the input modes. An asymmetric beam splitting ratio of the entangling beam splitter on the other hand reduces the amount of observable correlations along the anti--squeezed direction. The reason is that the contributions of the uncertainty originating from the anti--squeezing of the input modes cannot be cancelled simultaneously. In particular, for input squeezed states that are not minimum uncertainty states, the degree of correlations is reduced. This is a limitation for the application of the entanglement source in quantum information protocols, as the amount of useful entanglement is reduced. Furthermore it is desireable that the correlations in the conjugate observables have the same level. 

However, it is possible to observe quantum correlations also along a different direction which is more robust against asymmetries. Following the general definition of the quantum Stokes parameters in the dark plane from eqn.~(\ref{pol_sq_darkplane}), we can define the optimised observation direction $\hat S_\mathrm{opt}=\hat S(\theta_\mathrm{sq}-\gamma)=\sqrt{T}\hat S(\theta_\mathrm{sq})-\sqrt{R}\hat S(\theta_\mathrm{sq}+\pi/2)$, with $\cos\gamma=\sqrt{T}$ and $\sin\gamma=\sqrt{R}$, the corresponding orthogonal direction is given by $\hat S_{\mathrm{opt},\perp}=\hat S(\theta_\mathrm{sq}+\pi/2-\gamma)=\sqrt{R}\hat S(\theta_\mathrm{sq})+\sqrt{T}\hat S(\theta_\mathrm{sq}+\pi/2)$. The Stokes parameters along $\hat S_\mathrm{opt}$ and $\hat S_{\mathrm{opt},\perp}$ after the entangling beam splitter can be expressed in terms of the polarisation squeezed input modes
\begin{equation}\label{corr_opt}
\begin{array}{lclcl}
\hat S_\mathrm{opt,C}&=&\hat S_\mathrm{C}(\theta_\mathrm{sq}-\gamma)&=&\sqrt{T}\hat S_\mathrm{A}(\theta_\mathrm{sq})-\sqrt{R}\hat S_\mathrm{B}(\theta_\mathrm{sq}+\pi/2) \nonumber \\
\hat S_\mathrm{opt,\perp,C}&=&\hat S_\mathrm{C}(\theta_\mathrm{sq}+\pi/2-\gamma)&=&\sqrt{R}\hat S_\mathrm{B}(\theta_\mathrm{sq})+\sqrt{T}\hat S_\mathrm{A}(\theta_\mathrm{sq}+\pi/2) \nonumber \\
\hat S_\mathrm{opt,D}&=&\hat S_\mathrm{D}(\theta_\mathrm{sq}-\gamma)&=&\sqrt{T}\hat S_\mathrm{B}(\theta_\mathrm{sq})-\sqrt{R}\hat S_\mathrm{A}(\theta_\mathrm{sq}+\pi/2) \nonumber \\
\hat S_\mathrm{opt,\perp,D}&=&\hat S_\mathrm{D}(\theta_\mathrm{sq}+\pi/2-\gamma)&=&\sqrt{R}\hat S_\mathrm{A}(\theta_\mathrm{sq})+\sqrt{T}\hat S_\mathrm{B}(\theta_\mathrm{sq}+\pi/2). 
\end{array}
\end{equation}
From these relations it is evident that correlations of the type $g\hat S_\mathrm{opt,C}+\frac{1}{g}\hat S_{\mathrm{opt,\perp,D}}\rightarrow0$ and $\frac{1}{g}\hat S_{\mathrm{opt},\perp,C}+g\hat S_{\mathrm{opt,D}}\rightarrow0$ occur. A variable gain $g$ has been included to optimise the correlations. For a symmetric beam splitter, $g=1$, otherwise the correlations are optimised by $g=((TV_\mathrm{sq}+RV_\mathrm{asq})/(TV_\mathrm{asq}+RV_\mathrm{sq}))^{1/4}$, which depends on the beam splitting ratio and the degree of squeezing and the anti--squeezing of the input states, i.~e.\ their purity. Phyically speaking, the uncertainty areas of the two entangled output beams are deformed to ellipses if the beam splitter is asymmetric. The directions of $\hat S_\mathrm{obs}$ and $\hat S_\mathrm{obs,\perp}$ are oriented along the semi--major and the semi--minor axes of the ellipses. The application of the electronic gain is subsequently transforming the uncertainty ellipses into circles, it thus effectively accomplishes local squeezing operations.

The measurement of the correlations along this direction is optimal, i.~e. the maximum possible correlations can be observed according to the amount of entanglement that has been generated. In general, less entanglement is generated for more asymmetric beam splitting ratios, and also the excess noise in the anti--squeezed direction reduces the amount of possible entanglement if the beam splitter is not symmetric. We will present the results of the experimental characterisation for polarisation entanglement along the $\hat S(\theta_\mathrm{sq})$ and $\hat S(\theta_\mathrm{sq}+\pi/2)$ direction as well as the conjugate pair along $\hat S(\theta_\mathrm{sq}-\gamma)$ and $\hat S(\theta_\mathrm{sq}+\pi/2-\gamma)$.

\section{Experimental setup}

\begin{figure}[h] \begin{center}
\includegraphics[width=12cm]{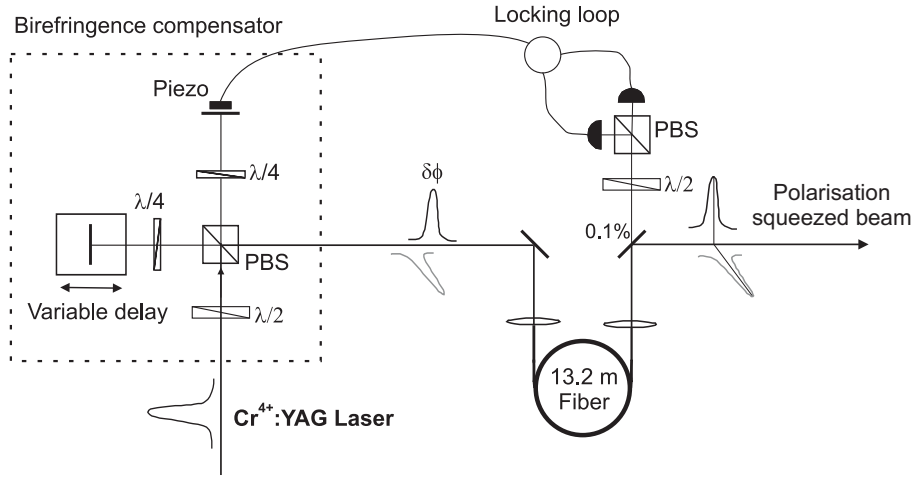}
\end{center}
\caption{Setup for the efficient generation of polarisation squeezing. A phase shift $\delta \phi$ between the two orthogonally polarised pulses is introduced before the fibre. Hence, a circularly polarised state is produced at the output. 0.1\% of the outgoing polarisation squeezed light is tapped off at a mirror and detected. The signal is used to control the relative phase between the polarisations and hence the polarisation state of the output mode via feedback. $\lambda/4$, $\lambda/2$: quarter--, half--wave plates, PBS: polarising beam splitter.}\label{polsqueezingsetup}
\end{figure}

In the experiment a Cr$^{4+}$:YAG laser with a central wavelength of 1497~nm was used. It produced soliton shaped pulses at a repetition rate of 163~MHz with a duration of 140~fs. These pulses were measured to be shot noise limited at our measurement frequency (17.5MHz) and thus can be assumed to be coherent. To produce polarisation squeezing we exploit an optimised setup based on the single pass of two orthogonally polarised light pulses through a birefringent fibre~\cite{heersink05.ol}, shown in Fig.~\ref{polsqueezingsetup}. Using 13.2~meters of fibre (3M FS-PM-7811, mode field diameter 5.7~$\mu$m, beat length 1.67~mm), two quadrature squeezed states were independently generated. 

These pulses were overlapped with a $\pi/2$ relative phase shift after the fibre. This was accomplished by using an active phase lock in the pre--compensation of the fibre birefringence which introduced a $\delta\phi$ relative shift between the two polarisation eigenmodes of the fibre (Fig.~\ref{polsqueezingsetup}). This resulted in a circularly polarised beam at the fibre output, mathematically described by $\langle \hat{S}_3 \rangle \neq 0$ and $\langle \hat{S}_1 \rangle = \langle \hat{S}_2 \rangle = 0$. The conjugate polarisation operators, which can exhibit polarisation squeezing, are then found in the plane given by $\hat{S}_1-\hat{S}_2$, referred to as the "dark plane". We derive our polarisation squeezing from Kerr squeezed states in which the squeezed quadrature is skewed by $\theta_\mathrm{sq}$ from the amplitude direction, where $\theta_\mathrm{sq}=0$ for amplitude squeezing. Thus the squeezed Stokes operator is given by $\hat{S}(\theta_\mathrm{sq})$ as defined in Eq.~(\ref{pol_sq_darkplane}). The orthogonal, anti--squeezed Stokes operator is $\hat{S}(\theta_\mathrm{sq}+\pi/2)$. We emphasise that these operators both have zero mean values. Furthermore, they both commute with the bright $\hat{S}_3$ component of the optical field.

The present setup has a number of advantages over previous fibre based experiments. Firstly, by employing the polarisation rather than the quadrature variables~\cite{silberhorn01.prl}, characterisation of all relevant parameters is possible by simple direct detection~\cite{korolkova02.pra}. In particular, the measurement of the optimised variables $\hat S(\theta_\mathrm{sq}-\gamma)$ and $\hat S(\theta_\mathrm{sq}+\pi/2-\gamma)$ is possible, which allows for the observation of the maximum correlations. Further, the single pass squeezing method allows us to avoid the intrinsic limitations of Sagnac loop squeezers and noticeably improved squeezing can be generated~\cite{heersink05.ol, corney06.prl}. This in turn allows for improved entanglement generation.

\begin{figure}[h]
\begin{center}
\includegraphics[width=12cm]{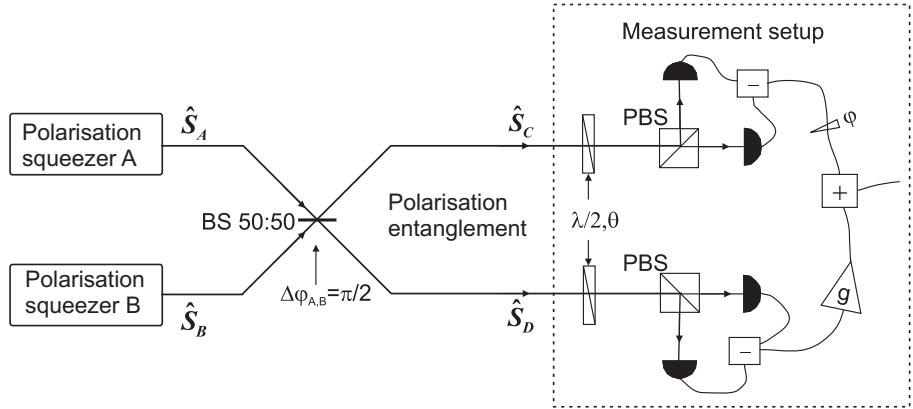} \end{center}
\caption{Setup for the generation of polarisation entanglement. Two polarisation squeezed beams interfere at a 50:50 beam splitter with a relative phase of $\Delta \varphi_\mathrm{A,B}=\pi/2$. In the two output ports, C~and~D, the dark plane Stokes parameters $\hat{S}_{C,D}(\theta_\mathrm{sq})$ and $\hat{S}_{C,D}(\theta_\mathrm{sq}+\pi/2)$ were measured. The photo--currents were added/subtracted to check for correlations. A variable gain g and phase shift $\varphi$ is introduced in the cables to minimise the variances. $\lambda/2$: half--wave plate, PBS: polarising beam splitter.}\label{polentsetup}
\end{figure}

Two such polarisation squeezed beams are simultaneously generated and are mixed on a 50:50~beam splitter (Fig.~\ref{polentsetup}). The two resulting intense beams, labeled C and D, are set via a phase lock to have equal intensity, i.~e.\ the two inputs are set to have a $\pi/2$ relative phase shift. The entangled outputs of the beam splitter thus are also circularly polarised. These beams are measured independently in two Stokes measurement apparatuses. These are optimised for measurements in the dark $\hat{S}_1-\hat{S}_2$ plane and thus are composed of only a half-wave plate ($\lambda/2$) followed by a polarising beam splitter (PBS). Appropriate rotation of the half--wave plate allows for the observation of the conjugate Stokes parameters which exhibit the entanglement, i.~e.\ the correlations in the squeezing and the anti--squeezing direction and in the optimised observation directions. The outputs of the PBS are detected by identical pairs of balanced photo--detectors based on custom made pin photo--diodes (98\% quantum efficiency at DC). The detection frequency of 17.5~MHz was chosen to avoid low frequency technical noise as well as the 163~MHz laser repetition rate, although in principle any frequency up to several THz is possible. The detected AC photocurrents are passively pairwise subtracted, added and monitored on a spectrum analyser (HP 8590E, 300~kHz resolution bandwidth, 30~Hz video bandwidth). 

\section{Results}

\begin{figure}[h]
\begin{center}
\includegraphics[width=14cm]{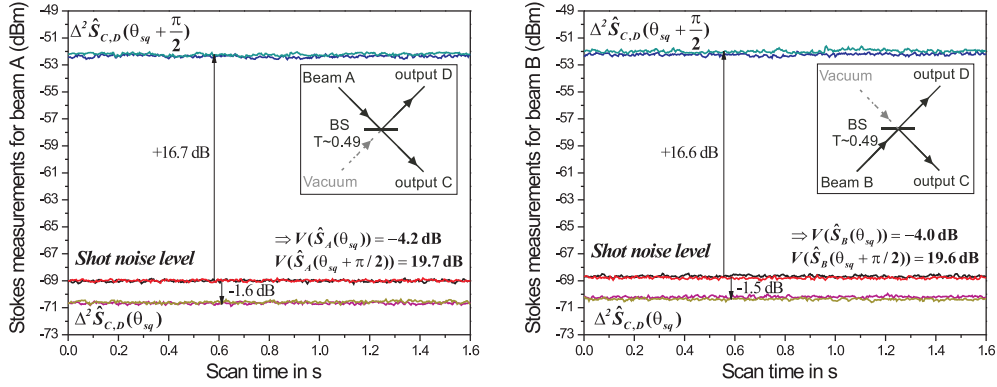}
\end{center}
\caption{Characterisation of the polarisation squeezing, located in the dark $\hat{S}_1-\hat{S}_2$ plane, for a total pulse power of 8.6~mW. From the measured squeezing levels of the output ports C and D, the squeezing levels of the input modes were inferred by taking into account the vacuum noise introduced by the beam splitter. The polarisation squeezing values for the input states A and B were \mbox{$\Delta^2_\mathrm{norm}\hat{S}_\mathrm{A}(\theta_\mathrm{sq})=-4.2$~dB}, \mbox{$\Delta^2_\mathrm{norm}\hat{S}_\mathrm{B}(\theta_\mathrm{sq})=-4.0$~dB}, \mbox{$\Delta^2_\mathrm{norm}\hat{S}_\mathrm{A}(\theta_\mathrm{sq}+\pi/2)=+19.7$~dB} and \mbox{$\Delta^2_\mathrm{norm}\hat{S}_\mathrm{B}(\theta_\mathrm{sq}+\pi/2)=+19.6$~dB}. The measurement frequency was at $17.5$~MHz, the resolution bandwidth was $300$~kHz and the resolution bandwidth was $30$~Hz. The electronic noise level, which was at $-84.5$~dBm was subtracted. }\label{polsq}
\end{figure}

In this section, we present the results of the characterisation of our entanglement source. Variances that were normalized to the respective mean values of the $\hat S_3$ parameter, which corresponds to the shot noise reference are denoted by $\Delta^2_\mathrm{norm}(\cdot)$. In a first step, the polarisation squeezing of the two input modes A and B was measured. We used the setup as depicted in Fig.~\ref{polentsetup}. In order to characterise the squeezing, we blocked the input modes A or B respectively and measured the polarisation squeezing of the output modes C and D. Each output mode showed reduced squeezing due to the vacuum fluctuations entering at the beam splitter. From the observed level of squeezing, which is presented in Fig.~\ref{polsq}, we can infer the amount of squeezing in the input modes. Polarisation squeezing of -4.2$\pm$0.3~dB was observed for the $\hat{S}_\mathrm{A}(\theta_\mathrm{sq})$ parameter of source A. Its canonic conjugate, $\hat{S}_\mathrm{A}(\theta_\mathrm{sq}+\pi/2)$, was anti--squeezed by +19.7$\pm$0.3~dB. The second beam exhibited similar squeezing levels of  -4.0$\pm$0.3~dB in $\hat{S}_\mathrm{B}(\theta_\mathrm{sq})$ and of +19.6$\pm$0.3~dB in $\hat{S}_\mathrm{B}(\theta_\mathrm{sq}+\pi/2)$. These noise traces as well as those for the polarisation entanglement were corrected for an electronic noise. The individual squeezed beams A and B exhibited a total optical power of 8.6~mW, corresponding to an energy of 53~pJ per pulse. The squeezing angle $\theta_\mathrm{sq}$ was around 4.5$^\circ$.

In the polarisation entanglement generation (Fig.~\ref{polentsetup}), the interference visibility between the squeezed outputs A and B was $>$98\%. Our polarisation entangled state was set to have a single nonzero Stokes parameter, namely $\hat{S}_3$. We used Eq.~(\ref{PeresHorodecki_productform}) to check for the non--separability of our output state. Non--classical correlations in the conjugate Stokes operators were observed along the $\hat S(\theta_\mathrm{sq})$ and $\hat S(\theta_\mathrm{sq}+\pi/2)$ directions and along $\hat S(\theta_\mathrm{sq}-\gamma)$ and $\hat S(\theta_\mathrm{sq}+\pi/2-\gamma)$ by measuring the respective Stokes parameters at the two output ports of the beam splitter and taking the variance of the sum and the difference signals. 

\begin{figure}[h]
\begin{center}
\includegraphics[width=14cm]{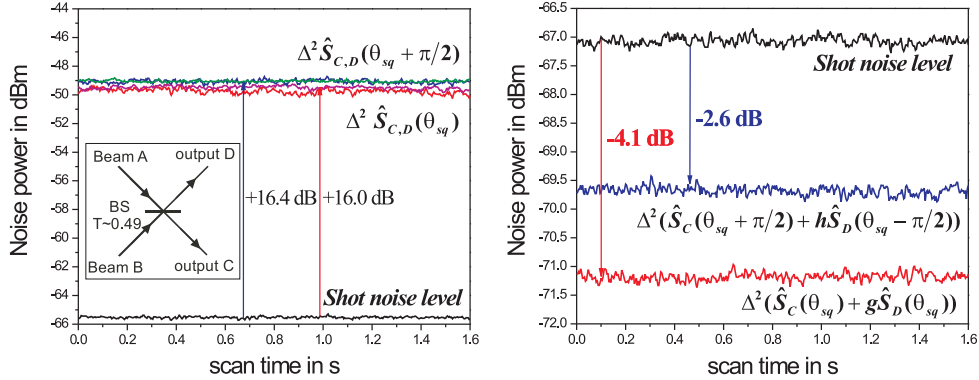}
\end{center}
\caption{Measurement of the noise of the entangled beam pair along $\hat S(\theta_\mathrm{sq})$ and $\hat S(\theta_\mathrm{sq}+\pi/2)$. The noise of the individual beams $\hat S_\mathrm{C,D}(\theta_\mathrm{sq})$ and $\hat S_\mathrm{C,D}(\theta_\mathrm{sq}+\pi/2)$ is plotted on the left side, the correlations $\Delta^2(\hat{S}_\mathrm{C}(\theta_\mathrm{sq}) + g\hat{S}_\mathrm{D}(\theta_\mathrm{sq}))$ and $\Delta^2(\hat{S}_\mathrm{C}(\theta_\mathrm{sq}+\pi/2) + h\hat{S}_\mathrm{D}(\theta_\mathrm{sq}-\pi/2))$ are plotted on the right side. Note the difference in the level of correlations of the two signals, which is a consequence of the asymmetric splitting ratio of the entangling beam splitter together with the high level of excess noise. The measurement frequency was at $17.5$~MHz, the resolution bandwidth was $300$~kHz and the resolution bandwidth was $30$~Hz. The electronic noise level, which was at $-85.5$~dBm was subtracted.}\label{polent_sq}
\end{figure}

In Fig.~\ref{polent_sq} $\Delta^2(\hat{S}_\mathrm{C}(\theta_\mathrm{sq})+g\hat{S}_\mathrm{D}(\theta_\mathrm{sq}))$ is plotted as well as the variances of the Stokes parameters of the individual modes at the output ports C and D and the corresponding shot noise level. We have optimised the parameter $g$ to minimise the variances of the correlation signals. Each individual mode is seen to exhibit a large excess noise (around $16$ dB of have been measured), as is typical for entangled states employing fibre based squeezers~\cite{silberhorn01.prl, gloeckl03.job}. Non--classical correlations in the $\hat{S}(\theta_\mathrm{sq})$ parameters are observed. We found \mbox{$\Delta^2_\mathrm{norm}(\hat{S}_\mathrm{C}(\theta_\mathrm{sq}) + g\hat{S}_\mathrm{D}(\theta_\mathrm{sq})) = 0.39\pm0.03$}, or -4.1$\pm$0.3~dB below the shot noise level. The variable gain $g$ in the detection system was necessary to balance slight variations in the detector coupling, gain and losses in the electronic subtractions, i.e.\ to set the gain of both detection setups nominally to unity. Experimentally this took the form of a relative variable attenuation ($g=0.91$ or 0.4~dB). Further, the relative phase of the electronic signals, $\varphi$ in Fig.~\ref{polentsetup}, was optimised by adjusting the relative cable lengths ($\varphi=0.51$~rad corresponding to a cable length difference of 0.92~m) such that maximal correlations were observed. This was mainly for the compensation of different optical pathlengths of the entangled light beams before the detectors. Both $g$ and $\varphi$ were set once before taking the measurements. 

The noise traces of the conjugate $\hat{S}(\theta_\mathrm{sq}+\pi/2)$ parameter are similar (see also Fig.~\ref{polent_sq}), however, less correlations were observed. Each individual signal exhibits a similarly high degree of noise. The correct combination of these signals highlights their strong correlation. To eliminate electronic offsets we once again took the sum signal, but introduced a $\pi$ relative phase shift between the measurement setups A and B by appropriate rotation of the half-wave plates, i.e.\ \mbox{$\hat S(\theta_\mathrm{sq}+\pi/2)\rightarrow-\hat S(\theta_\mathrm{sq}-\pi/2)$}. The correlated signal is thus given by \mbox{$\hat{S}_\mathrm{C}(\theta_\mathrm{sq}+\pi/2)+h\hat{S}_\mathrm{D}(\theta_\mathrm{sq}-\pi/2)$}. The measurements confirm the correlations, and the variance is $\Delta^2_\mathrm{norm}(\hat{S}_\mathrm{C}(\theta_\mathrm{sq}+\pi/2) +h\hat{S}_\mathrm{D}(\theta_\mathrm{sq}-\pi/2))=0.55\pm0.03$ or -2.6$\pm$0.3~dB below the shot noise level. The parameter $h$ was also optimised for this measurement, but had the same value as $g$ in the measurement above.
\begin{figure}[h]
\begin{center}
\includegraphics[width=14cm]{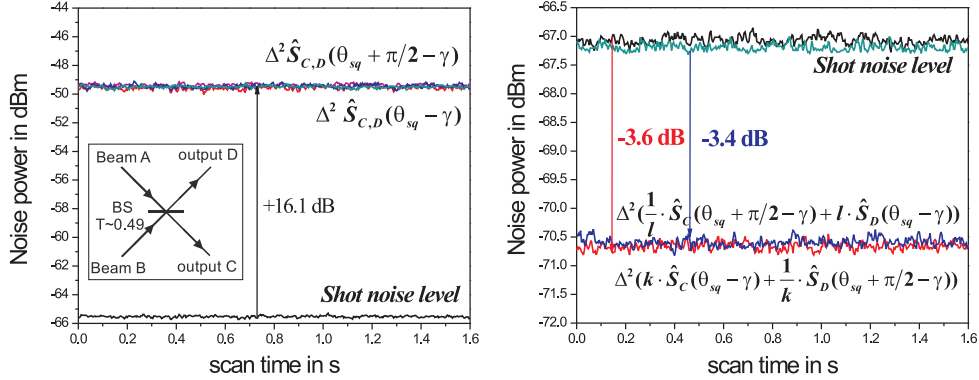}
\end{center}
\caption{Measurement of the noise of the entangled beam pair along $\hat S(\theta_\mathrm{sq}-\gamma)$ and $\hat S(\theta_\mathrm{sq}+\pi/2-\gamma)$. The noise of the individual beams $\hat S_\mathrm{C,D}(\theta_\mathrm{sq}-\gamma)$ and $\hat S_\mathrm{C,D}(\theta_\mathrm{sq}+\pi/2-\gamma)$ is plotted on the left side, the correlations $\Delta^2(k\hat{S}_\mathrm{C}(\theta_\mathrm{sq}-\gamma) + \frac{1}{k}\hat{S}_\mathrm{D}(\theta_\mathrm{sq}+\pi/2-\gamma))$ and $\Delta^2(\frac{1}{l}\hat{S}_\mathrm{C}(\theta_\mathrm{sq}+\pi/2-\gamma) + l\hat{S}_\mathrm{D}(\theta_\mathrm{sq}-\gamma))$ are plotted on the right side. The angle $\gamma$ is given by the beam splitting ratio and is approximately at $45^\circ$.
The level of the correaltions is almost identical. The measurement frequency was at $17.5$~MHz, the resolution bandwidth was $300$~kHz and the resolution bandwidth was $30$~Hz. The electronic noise level, which was at $-85.5$~dBm was subtracted.}\label{polent_45}
\end{figure}
The application of the non--separability criterion of Eq.~(\ref{PeresHorodecki_productform})
\begin{eqnarray}
\sqrt{\Delta^2_\mathrm{norm}(\hat{S}_\mathrm{C}(\theta_\mathrm{sq})+g\hat{S}_\mathrm{D}(\theta_\mathrm{sq})) \cdot \Delta^2_\mathrm{norm}(\hat{S}_\mathrm{C}(\theta_\mathrm{sq}+\pi/2)+h\hat{S}_\mathrm{D}(\theta_\mathrm{sq}-\pi/2))}\nonumber \\ 
\qquad = \sqrt{0.39 \cdot 0.55} = 0.46\pm0.03< 1,\label{entanglementgemessen} 
\end{eqnarray}
proves that a highly correlated non--separable quantum state in the Stokes variables has been generated.

In the next step, we verified the correlations along $\hat S_\mathrm{obs}=\hat S(\theta_\mathrm{sq}-\gamma)$ and $\hat S_\mathrm{obs,\perp}=\hat S(\theta_\mathrm{sq}+\pi/2-\gamma)$. For the measurement, we had a beam splitting ratio which was close to 50:50, as a result, $\gamma$ had to be chosen to be approximately $\pi/4$. The noise level of the individual modes was again well above the shot noise level (around $16.1$ dB). The correlations are characterised in terms of $\Delta^2(k\hat S_\mathrm{C}(\theta_\mathrm{sq}-\gamma)+\frac{1}{k}\hat S_\mathrm{D}(\theta_\mathrm{sq}+\pi/2-\gamma))$ and $\Delta^2(\frac{1}{l}\hat S_\mathrm{C}(\theta_\mathrm{sq}+\pi/2-\gamma)+l\hat S_\mathrm{D}(\theta_\mathrm{sq}-\gamma))$.  The results are plotted in Fig \ref{polent_45}. We found $\Delta^2_\mathrm{norm}(k\hat S_\mathrm{C}(\theta_\mathrm{sq}-\gamma)+\frac{1}{k}\hat S_\mathrm{D}(\theta_\mathrm{sq}+\pi/2-\gamma))=0.44\pm0.03$, or $-3.6$dB below the shot noise level. The gain factor $k$ was optimised to maximise the correlations, furthermore, the rotation of the $\lambda/2$--wave plate to define the angle $\gamma$ was fine adjusted. The correlations of the respective conjugate parameters were similar, we measured $\Delta^2_\mathrm{norm}(\frac{1}{l}\hat S_\mathrm{C}(\theta_\mathrm{sq}+\pi/2-\gamma)+l\hat S_\mathrm{D}(\theta_\mathrm{sq}-\gamma))=0.46\pm0.03$, or $-3.4$dB below the shot noise level. These measurements again show the non--separability of our state as we have 
\begin{equation}
\sqrt{0.44 \cdot 0.46} = 0.45\pm0.03< 1. 
\end{equation}
It can be seen that the correlations along $\hat S(\theta_\mathrm{sq}-\gamma)$ and $\hat S(\theta_\mathrm{sq}+\pi/2-\gamma)$ is more robust against beam splitter asymmetries, and the correlations of the conjugate pair are more symmetric making the resource more useful for applications. 

Let us summarise the influence of the beam splitting ratio of the entangling beam splitter on the degree of observable correlations and the amount of entanglement that is generated. In the precence of our states' large excess noise, which is 20~dB or more above that of a minimum uncertainty state, the correlations can be optimised using the appropriate measurement strategy. As it is evident from equations (\ref{eq_pol_ent_bs}), in the presence of large excess noise, the correlations along the squeezed and anti--squeezed direction cannot be minimised simultaneously, as the states' excess noise does not cancel. This explains the different results for the degree of quantum correlations in our measurements. However, measuring along the Stokes parameters as described in equations (\ref{corr_opt}), i.~e.\ measurement along $\hat S(\theta_\mathrm{sq}-\gamma)$ and $\hat S(\theta_\mathrm{sq}+\pi/2-\gamma)$ allows for the detection of the maximum possible correlations. The level of observed correlations is the same for both combinations of the conjugate variables. 
The amount of observable correlations in that case are given by the degree of entanglement generated in the system. The amount of entanglement is governed by the beam splitting ratio as well as the purity of the system, if the beam splitting ratio is asymmetric.

\section{Conclusions and outlook}

We have demonstrated the efficient production of a state exhibiting strong quantum correlations in the optical polarisation variables. Our squeezing source is relatively simple and robust as few locking loops are required. Exploiting the Stokes parameters, the entanglement was easily measured in simple direct detection in contrast to other experiments using intense beams and quadrature variables~\cite{ou92.prl, silberhorn01.prl}.

One limiting factor observed in our experiments was the inherent asymmetry of the entangling beam splitter in combination with our states' large excess noise when the correlations were measured in the Stokes parameters which were oriented in the squeezing and anti--squeezing direction of the input beams. However, by measuring in a properly chosen pair of conjugate Stokes parameters, we observed correlations which were robust against the beam splitting ratio of the entangling beam splitter. Using this measurement strategy, our states can be called polarisation entangled according to the EPR--criterion \cite{reid89.pra} even in the presence of the large excess noise that limits the minimisation of the conditional variances.

As the amount of entanglement that is generated by interfering two polarisation squeezed beams depends on the input states' purity (for an asymmetric beam splitting ratio), it is worthwhile to examine methods to reduce the excess noise from fibre squeezers. These include purification schemes~\cite{gloeckl06.prl} for the squeezing resource, which could be extended to polarisation variables and the use of photonic crystal fibres~\cite{elser06.prl}. 
The entanglement source can then potentially be used for the quantum key distribution scheme described im Ref. \cite{cry.prl}. Further, the simulation work of Corney~\textit{et al.}~\cite{corney06.prl} could be extended to determine the fibre length and or traits exhibiting the best trade--off in terms of squeezing--excess noise. A deeper understanding and thorough characterisation of the polarisation states, a first step of which has been taken in Ref.~\cite{marquardt07.xxx}, could also lead to improved results. 

Our source for polarisation entangled states is suited for applications in quantum information and communication, particularly due to the ease of detection without the need for an external phase reference beam. For example, the extension of our work on distillation of quantum states afflicted by non--Gaussian noise~\cite{heersink06.prl} to entangled states is of interest. 

\ack
We thank M. Chekhova for help and useful discussions. This work was supported by the EU project COVAQIAL (Project No. FP6-511004).

\section*{References}


\end{document}